%% file: main.tex
\documentclass[conference]{IEEEtran}
\usepackage[utf8]{inputenc}
\usepackage[T1]{fontenc}
\usepackage{graphicx}
\usepackage{tabularx,booktabs}
\usepackage{amsmath,amssymb,amsthm,mathtools}
\usepackage{cite}
\usepackage{xcolor}
\usepackage{adjustbox}
\usepackage{gensymb}
\usepackage[caption=false,font=footnotesize]{subfig}
\usepackage{listings}
\usepackage{stfloats} 
\usepackage{wasysym} 
\usepackage{pgfplots} 
\pgfplotsset{compat=newest}
\usetikzlibrary{patterns}
\usepackage{tikz}

\newtheorem{lemma}{\textbf{Lemma}}

\newtheorem{theorem}{\textbf{Theorem}}

\title{Opportunistic Cooperative Sensing in mmWave ISAC Networks: A Meta-Distribution Analysis}

\author{
\IEEEauthorblockN{
Yasser Nabil\IEEEauthorrefmark{1},
Hesham ElSawy\IEEEauthorrefmark{2},
and Hossam S. Hassanein\IEEEauthorrefmark{2}
}

\IEEEauthorblockA{
\IEEEauthorrefmark{1}Department of Electrical and Computer Engineering, Queen's University, Kingston, ON, Canada
}

\IEEEauthorblockA{
\IEEEauthorrefmark{2}School of Computing, Queen's University, Kingston, ON, Canada
}
}

\begin{document}

\maketitle

\begin{abstract}
We propose an opportunistic cooperative sensing architecture for millimeter-wave (mmWave) integrated sensing and communication (ISAC) networks that harvests bistatic echoes from already-active transmissions of neighboring base stations (BSs). Each target is sensed by its nearest BS, while neighboring BSs contribute only when their scheduled beams opportunistically illuminate the target. Cooperation requires neither beam reconfiguration nor dedicated air-interface resources and can leverage existing 5G transport and edge-computing infrastructure for synchronization and local echo processing. Using stochastic geometry, we characterize the sensing meta-distribution for cooperative ISAC, quantifying target-level reliability beyond spatial averages when multiple cooperative echoes are combined, while also analyzing sensing and communication coverage and ergodic rates. Results show that opportunistic cooperation substantially improves average sensing performance and target-level reliability. For instance, at a sensing signal-to-interference-plus-noise ratio (SINR) threshold of $5$~dB, increasing the maximum cooperative-cluster size from $1$ to $6$ raises the percentage of targets whose conditional sensing success probability exceeds $90\%$ by over $55$ percentage points. The results further reveal practical design tradeoffs in cluster sizing, BS densification, and sensing--communication energy allocation.
\end{abstract}


\IEEEpeerreviewmaketitle

\section{Introduction}

Integrated sensing and communication (ISAC) is envisioned as an important capability for reliability-sensitive 6G applications, including autonomous driving, unmanned aerial vehicle (UAV) navigation, and industrial automation, where stringent sensing reliability must coexist with high-rate connectivity \cite{liu2022integrated,zhang2026integrated}. At millimeter-wave (mmWave) frequencies, however, reliable sensing is challenged by severe two-way propagation loss, intermittent line-of-sight (LoS) availability due to blockage, strong co-channel interference in dense deployments, and residual self-interference (SI) under full-duplex (FD) operation \cite{olson2023coverage,11270002}. These effects induce high spatial variability in sensing performance and may lead to sensing blind spots.

To mitigate these limitations, cooperative ISAC enables multiple base stations (BSs) to sense the same target from different viewing geometries, improving robustness to blockage and exploiting viewing-angle-dependent radar cross-section (RCS) diversity \cite{11270002,10769538}. Realizing these gains, however, often entails sensing-aware transmission design and coordination. For instance, multi-BS schemes optimize transmit beamforming for sensing objectives \cite{yang2024coordinated}, cell-free architectures employ centralized processing and sensing-aware power allocation \cite{behdad2024multi}, and coordinated multipoint schemes exploit bistatic reflections through jointly optimized precoders and inter-BS sharing of channel state information and user data \cite{babu2024precoding}. These approaches therefore rely on cooperation-specific transmission strategies and/or centralized processing, with some requiring inter-BS data or sensing-observation exchange. This motivates a complementary question: can cooperative sensing gains be harvested from already-active ISAC transmissions without requiring sensing-specific beam scheduling at neighboring BSs, reserving additional air-interface resources for cooperation, or requiring centralized sensing-data fusion?

Motivated by this, we propose an opportunistic cooperative sensing architecture for dense mmWave ISAC networks that harvests multi-BS transmit diversity from already-active neighboring-BS transmissions. Each target is sensed by its nearest BS, while neighboring BSs retain their local scheduling and contribute only when one of their active beams illuminates the target. The sensing BS combines available bistatic echoes with its monostatic echo, so the monostatic baseline is always retained while additional illuminators provide spatial diversity. Cooperation neither redirects nor reserves neighboring-BS beams, requires no dedicated air-interface resources for cooperation, and avoids inter-BS exchange of raw sensing observations, relying on timing/synchronization and beam-scheduling metadata exchanged over existing 5G transport infrastructure and local echo processing at the sensing BS.

Because the availability of cooperative illuminators varies across network realizations, assessing sensing reliability requires characterizing not only average performance but also the fraction of targets meeting stringent reliability requirements. The sensing meta-distribution provides this target-level reliability view by characterizing the distribution of the conditional sensing success probability across targets \cite{7345601,zhu2026meta}. Specifically, $\bar{F}_p(t)$ gives the fraction of targets whose conditional sensing success probability exceeds $t$; for example, $\bar{F}_p(0.9)=0.75$ means that $75\%$ of targets have a conditional sensing success probability exceeding $0.9$ at a given sensing signal-to-interference-plus-noise ratio (SINR) threshold. Recent works studied sensing meta-distributions for automotive radar \cite{shah2025fine}, bistatic joint radar-communication \cite{ram2022optimization}, and monostatic ISAC networks \cite{zhu2026meta}. However, these studies do not characterize the meta-distribution when multiple distinct cooperative echoes from different BSs are jointly combined for the same target.
To this end, the main contributions are:
\begin{itemize}

\item We propose a novel opportunistic ISAC architecture that harvests multistatic echoes from already-active neighboring-BS transmissions without beam redesign or air-interface resource reservation for cooperation, with all echo processing performed locally at the sensing BS.

\item We develop a tractable stochastic-geometry framework that captures opportunistic BS participation through Bernoulli-gated echo combining under co-channel and target-reflected interference, enabling joint sensing and communication performance analysis.

\item To the best of our knowledge, we provide the first sensing meta-distribution analysis for cooperative ISAC with combined multi-BS echoes, quantifying target-level reliability and the fraction of targets meeting stringent sensing requirements beyond spatial averages.
\end{itemize}

\section{System Model}

Consider a large mmWave network in which ISAC BS locations follow a Poisson point process (PPP) $\Phi_{\mathrm{BS}}$ with density $\lambda_{\mathrm{BS}}$. The network simultaneously senses targets and serves mobile users (MUs). Each BS forms $N_b$ simultaneous non-overlapping directional beams under a spatial-division multiple-access (SDMA) architecture, where $N_b$ is a design parameter. Every beam uses the full ISAC bandwidth $W$, reused across BSs, and has maximum gain $G_m$ and $3$-dB beamwidth $\theta_b$, while MUs use omnidirectional antennas. Each beam senses a target and, owing to the high MU density, simultaneously serves a scheduled MU in the same direction.\footnote{Using the full bandwidth $W$ on each beam provides fine range resolution and can help resolve multiple targets within a beam. Multi-target sensing is left for future work; the present analysis focuses on typical-target reliability.}

\begin{figure}[t]
\centering
\includegraphics[width=0.31\textwidth]{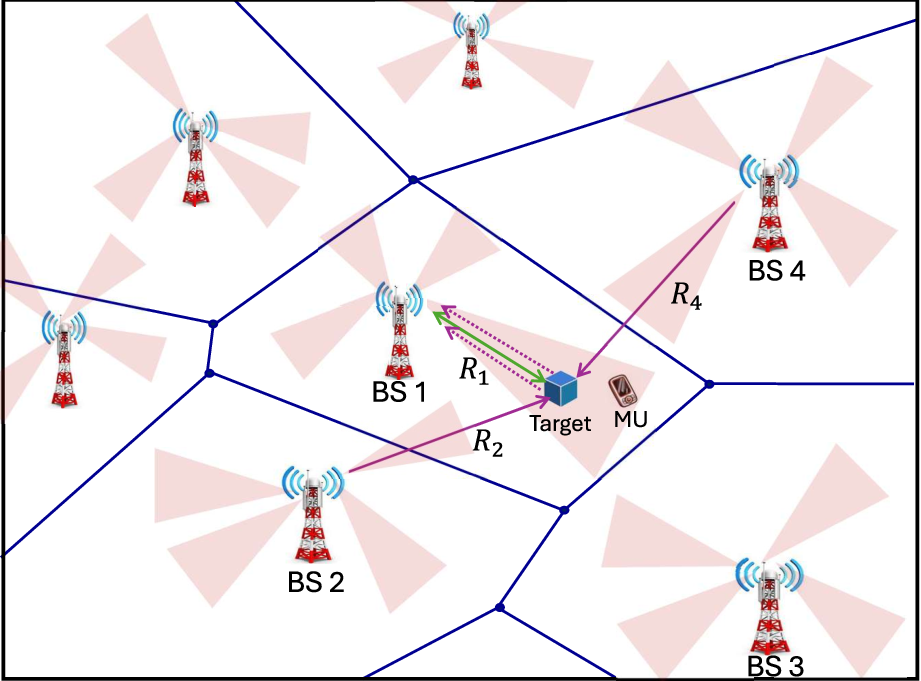}
\caption{Illustration of opportunistic cooperative ISAC with $N_c=4$. BS~1 receives its monostatic echo and opportunistic bistatic echoes from BS~2 and BS~4; BS~3 does not contribute because none of its beams illuminates the typical target. Here, $R_1$ and $R_n$, $2\le n\le N_c$, denote the target-to-sensing-BS and target-to-$n$-th-nearest-BS distances, respectively.}
\label{sys_mod}
\end{figure}

To enhance sensing reliability, an opportunistic cooperative cluster is formed by the $N_c$ nearest BSs to each target, with $N_c$ denoting the maximum cluster size. The nearest BS is the sensing BS and provides the monostatic echo, while the remaining $N_c-1$ BSs are candidate BSs. Because candidate-BS beam directions follow local ISAC scheduling and may vary across slots, target illumination is probabilistic. That is, a candidate BS contributes a bistatic echo only when one of its already-active beams illuminates the target over a LoS path, as illustrated in Fig.~\ref{sys_mod}. Thus, the number of echo contributors is a random variable in $[1,N_c]$, governed by the BS deployment geometry, blockage, and local beam scheduling.

\setlength{\textfloatsep}{5pt}

\begin{figure}[t]
\centering
\includegraphics[width=0.302\textwidth]{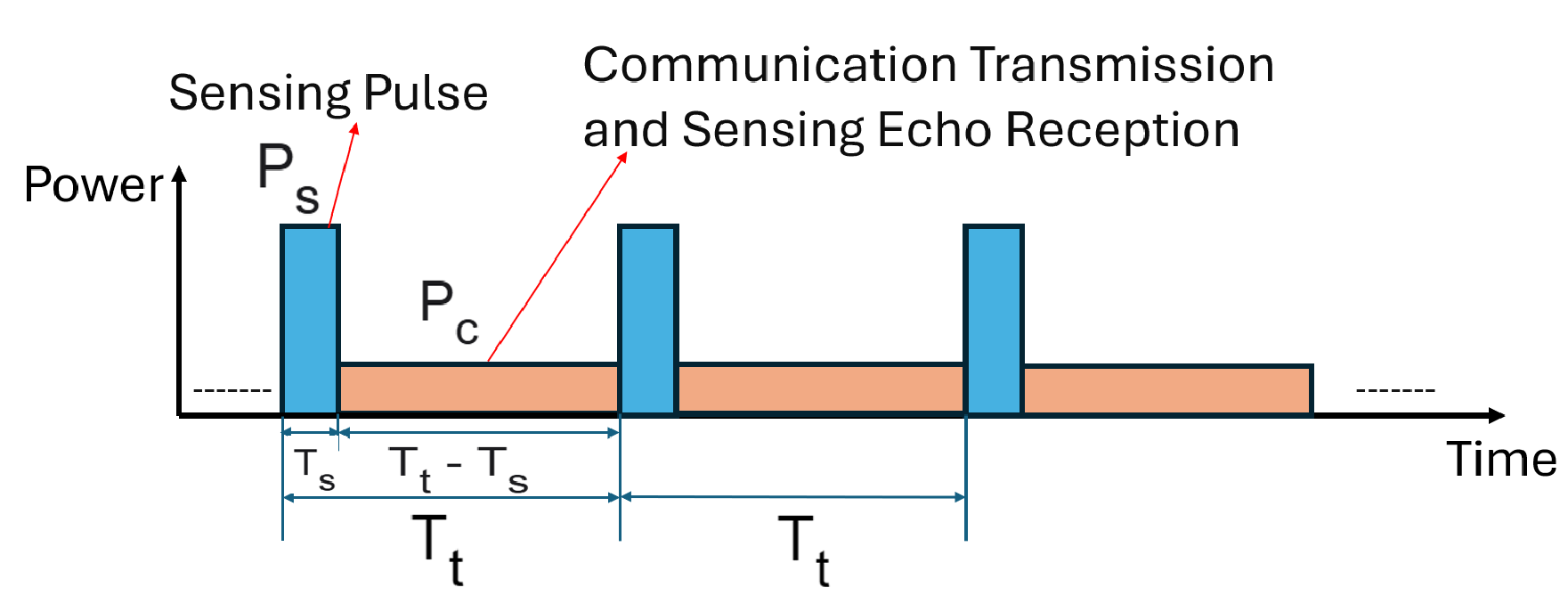}
\caption{An illustration of the unified ISAC signal.}
\label{sgn_mod}
\end{figure}

All BSs use an ISAC signal structure known network-wide. We adopt the unified signal in \cite{xiao2022waveform} for simultaneous sensing and communication. Each slot of duration $T_t$ comprises a sensing pulse of width $T_s$, followed by a simultaneous downlink-transmission/echo-reception interval of duration $T_t-T_s$, as shown in Fig.~\ref{sgn_mod}. We set $T_s=1/W$ and $T_t=2R_{\max}/c$, where $R_{\max}$ is the maximum unambiguous sensing range and $c$ is the speed of light. With energy budget $E$ per beam per slot and splitting factor $0\le\gamma\le1$, energies $\gamma E$ and $(1-\gamma)E$ are allocated to sensing and communication, respectively, yielding powers $P_s=\gamma E/T_s$ and $P_c=(1-\gamma)E/(T_t-T_s)$.

Downlink transmission and echo reception share the same beam, requiring self-interference cancellation (SIC) at the sensing BS. Under imperfect SIC \cite{xiao2022waveform}, let $\zeta_{\mathrm{SI}}$ denote the fraction of $P_c$ remaining as residual SI. Unlike conventional time-division sensing/communication, the ISAC signal uses a short pulse to initiate sensing and the remaining slot for simultaneous downlink transmission and echo reception. Hence, sensing and communication are coupled through shared bandwidth, energy split $\gamma$, residual SI, and network interference.

Candidate BSs retain their local ISAC scheduling and provide only control metadata carrying timing/synchronization information and the directions of their active beams. Together with the known sensing waveform, BS locations, and a coarse target location estimate from prior sensing \cite{meng2024network,olson2023coverage,10769538}, this metadata enables the sensing BS to identify illuminating transmissions, associate bistatic echoes, and align timing. The metadata can be exchanged over existing 5G transport infrastructure, often via optical-fiber links \cite{liu2022integrated}\footnote{We assume sufficiently accurate inter-BS timing synchronization. Quantifying the impact of residual synchronization errors is left for future work.}, while local echo processing at the sensing BS can be supported by 5G multi-access edge computing (MEC) and its evolution toward 6G \cite{zhang2026integrated,11433077}. Cooperation therefore relies on control signaling and local processing, without requiring additional ISAC beams, dedicated time/frequency resources, or inter-BS exchange of raw transmit waveforms or sensing observations/results.

Following the distance-dependent mmWave blockage model in \cite{rebato2019stochastic}, $p_{\mathrm{LOS}}(r)=e^{-\eta_b r}$, where $\eta_b$ depends on blocker density and geometry. Consistent with dense-mmWave analysis \cite{yu2017coverage}, we assume the MU and the target have LoS links to their serving BS (BS~1 in Fig.~\ref{sys_mod}). In contrast, target links to other BSs, including candidate BSs in the cooperative cluster, and co-channel interference links may be LoS or non-line-of-sight (NLoS). For communication and direct co-channel interference links in state $\chi\in\{L,N\}$, where $L$ and $N$ denote LoS and NLoS, the channel power gain is $C_\chi r^{-\alpha_\chi}g_\chi$, where $C_\chi$ and $\alpha_\chi$ are the path-loss intercept and exponent, and $g_\chi$ is a unit-mean Nakagami-$m$ Gamma fading gain with parameter $m_\chi$.
Target RCS fluctuations follow the Swerling-I model, with monostatic and bistatic RCS terms modeled as independent exponential random variables \cite{behdad2024multi}, a standard model widely used in system-level ISAC analyses \cite{olson2023coverage,10769538,11270002} for fluctuating targets.

\section{Analytical Framework}

\subsection{Cooperative Sensing Analysis}

We analyze the sensing reliability of a typical target. The sensing BS combines its monostatic echo with the available bistatic echoes from candidate BSs. To model a practically relevant sensing scenario, and in line with ISAC sensing models \cite{yang2024coordinated,olson2023coverage,zhu2026meta,10769538}, only direct LoS sensing echoes are treated as useful. Exploiting NLoS echoes would require knowledge of scatterer geometry for reliable path association and delay alignment, while at mmWave frequencies such echoes also incur higher propagation loss. We distinguish two interference types: (i) direct co-channel interference from other BSs received at the sensing BS over LoS or NLoS links; and (ii) target-reflected interference generated by out-of-cluster BSs whose beams illuminate the target over LoS and are reflected by the target toward the sensing BS.
\subsubsection{Aggregate Echo Power and Sensing SINR}

Following the monostatic and bistatic range equations, the echo powers are
\par
{\small
\begin{equation}
P_{\mathrm{mono}}
=\frac{P_sG_m^2\lambda^2\sigma_m}
{(4\pi)^3R_1^{2\alpha_L}},
\;\;
P_{\mathrm{bi},n}
=\frac{P_sG_m^2\lambda^2\sigma_{b,n}}
{(4\pi)^3R_1^{\alpha_L}R_n^{\alpha_L}},
\; n=2,\dots,N_c.
\end{equation}
}
Here, $\lambda$ is the wavelength. For simplicity, the sensing beam is modeled with constant gain $G_m$ within the $3$-dB beamwidth and zero gain outside. The monostatic and bistatic RCSs, $\sigma_m$ and $\sigma_{b,n}$, respectively, are exponential with means $\sigma_{\mathrm{av,mono}}$ and $\sigma_{\mathrm{av,bi}}$, so the corresponding echo powers are also exponential with means
\begin{equation}
\mu_{\mathrm{mono}}
=\frac{P_sG_m^2\lambda^2\sigma_{\mathrm{av,mono}}}
{(4\pi)^3R_1^{2\alpha_L}},
\quad
\mu_{\mathrm{bi},n}
=\frac{P_sG_m^2\lambda^2\sigma_{\mathrm{av,bi}}}
{(4\pi)^3R_1^{\alpha_L}R_n^{\alpha_L}}.
\end{equation}

We model the combined useful sensing power as the sum of
monostatic and bistatic echo powers \cite{10747175}:
\begin{equation}
\label{eq:Ptot_def}
P_{\mathrm{tot}}
=
P_{\mathrm{mono}}
+\sum_{n=2}^{N_c}
\mathbf{1}_{\{C_n\}}P_{\mathrm{bi},n}.
\end{equation}
Here, $C_n$ denotes the event that candidate BS~$n$ illuminates the target over a LoS link with one of its already-active beams, and $\mathbf{1}_{\{C_n\}}$ is an indicator function. Under random beam orientation, the probability that one of the $N_b$ active beams covers the target direction is $p_b\triangleq N_b\theta_b/(2\pi)$. Since beam scheduling and blockage are conditionally independent given $\Phi_{\mathrm{BS}}$, the conditional participation probability is
\begin{equation}
p_n\triangleq\mathbb{P}(C_n\mid\Phi_{\mathrm{BS}})
=p_{\mathrm{LOS}}(R_n)p_b.
\end{equation}

Conditioned on $\Phi_{\mathrm{BS}}$, candidate-BS contributions are independent. Hence, the mean and variance of $P_{\mathrm{tot}}$ are
\par
{\small
\begin{equation}
\label{eq:mean_var_Ptot}
\begin{aligned}
\mathbb{E}[P_{\mathrm{tot}}\mid\Phi_{\mathrm{BS}}]
&=\mu_{\mathrm{mono}}+\sum_{n=2}^{N_c}p_n\mu_{\mathrm{bi},n},\\
\mathrm{Var}(P_{\mathrm{tot}}\mid\Phi_{\mathrm{BS}})
&=\mu_{\mathrm{mono}}^2+\sum_{n=2}^{N_c}p_n(2-p_n)\mu_{\mathrm{bi},n}^2.
\end{aligned}
\end{equation}
}
\noindent
The variance follows from the independence of the candidate contributions and the Bernoulli-gated exponential form of each bistatic term $\mathbf{1}_{\{C_n\}}P_{\mathrm{bi},n}$. For analytical tractability, we use moment matching to approximate the conditional distribution of $P_{\mathrm{tot}}$ by a Gamma distribution:
$P_{\mathrm{tot}}\mid\Phi_{\mathrm{BS}}\approx\mathrm{Gamma}(k_{\mathrm{eff}},\theta_{\mathrm{eff}})$, with shape and scale parameters
\par
{\small
\begin{equation}
\label{eq:gamma_approx}
k_{\mathrm{eff}}
=\frac{\mathbb{E}[P_{\mathrm{tot}}\mid\Phi_{\mathrm{BS}}]^2}
{\mathrm{Var}(P_{\mathrm{tot}}\mid\Phi_{\mathrm{BS}})},
\qquad
\theta_{\mathrm{eff}}
=\frac{\mathrm{Var}(P_{\mathrm{tot}}\mid\Phi_{\mathrm{BS}})}
{\mathbb{E}[P_{\mathrm{tot}}\mid\Phi_{\mathrm{BS}}]}.
\end{equation}
}

The sensing SINR at the sensing BS is modeled as
\begin{equation}
\label{eq:SINR_S_compact}
\mathrm{SINR}_S
=
\frac{P_{\mathrm{tot}}}
{I_T+I_{L_S}+I_{N_S}+K_BTW+P_c\zeta_{\mathrm{SI}}},
\end{equation}
where $I_T$ is target-reflected interference, $I_{L_S}$ and $I_{N_S}$ are LoS and NLoS direct co-channel interference, $K_BTW$ is thermal-noise power with $K_B$ Boltzmann's constant and $T$ the receiver temperature, and $P_c\zeta_{\mathrm{SI}}$ is residual SI after cancellation.

Let $R_i$ denote the distance from BS~$i$ to the target. For each BS~$i\neq1$, let $r_i$ denote its distance to the sensing BS and $\vartheta_i$ the angular separation, as seen from the sensing BS, between the target direction and the direction to BS~$i$. Define 
$A_i\triangleq\mathbf{1}_{\{|\vartheta_i|\leq\theta_b/2\}}$, 
which indicates whether the direct signal from BS~$i$ falls within the sensing BS receive beam. Let $B_i$ indicate that one of the $N_b$ active beams of BS~$i$ points toward the sensing BS, such that 
$\mathbb{P}(B_i=1\mid\Phi_{\mathrm{BS}})=p_b$. Conditioned on $\Phi_{\mathrm{BS}}$, $A_i$ is fixed by the relative geometry, whereas $B_i$ remains a slot-dependent beam-scheduling variable.
Let $X_i\in\{L,N\}$ denote the propagation state of the link from BS~$i$ to the sensing BS, with
$\mathbb{P}(X_i=L\mid\Phi_{\mathrm{BS}})=p_{\mathrm{LOS}}(r_i)$
and $X_i=N$ otherwise. The direct interference components are
\par
{\small
\begin{equation}
\label{eq:ILS_INS_def}
I_{\chi_S}
=
P_cG_m^2C_\chi
\!\!\!\!\sum_{\mathrm{BS}_i\in\Phi_{\mathrm{BS}}\setminus\{\mathrm{BS}_1\}}\!\!\!\!
A_iB_i\mathbf{1}_{\{X_i=\chi\}}g_{\chi,i}r_i^{-\alpha_\chi},
\; \chi\in\{L,N\}.
\end{equation}
}
Target-reflected interference is generated by BSs outside the cooperative cluster whose active beams illuminate the target over LoS paths. Let $D_i\in\{0,1\}$ indicate this event, with
$\mathbb{P}(D_i=1\mid\Phi_{\mathrm{BS}})
=p_{\mathrm{LOS}}(R_i)p_b$. Then
\par
{\small
\begin{equation}
\label{eq:IT_def}
I_T
=
\frac{P_sG_m^2\lambda^2}{(4\pi)^3R_1^{\alpha_L}}
\sum_{\mathrm{BS}_i\in\Phi_{\mathrm{BS}},\,i>N_c}
D_i\sigma_{b,i}R_i^{-\alpha_L},
\end{equation}
}
where $\sigma_{b,i}$ follows the same exponential bistatic RCS model with mean $\sigma_{\mathrm{av,bi}}$.

\subsubsection{Conditional Cooperative Sensing Success Probability}

To characterize sensing reliability across network realizations, we condition on the BS geometry $\Phi_{\mathrm{BS}}$ and define the conditional cooperative sensing success probability as
\begin{equation}
\label{eq:cond_cov_def}
\mathcal{P}_{\mathrm{S}}(\Phi_{\mathrm{BS}};\tau)
\triangleq
\mathbb{P}\!\left(
\mathrm{SINR}_S>\tau
\,\middle|\,
\Phi_{\mathrm{BS}}
\right).
\end{equation}
Conditioned on $\Phi_{\mathrm{BS}}$, the remaining randomness arises from transmit-beam alignment, LoS/NLoS link states, fading, and RCS fluctuations. We first derive the conditional Laplace transforms (LTs) of the interference components.

\begin{lemma}
\label{lem:direct_LT}
Conditioned on $\Phi_{\mathrm{BS}}$, the LT of the aggregate direct co-channel interference is
\par
{\small
\begin{equation}
\label{eq:LT_direct_cond}
\mathcal{L}_{I_{L_S}+I_{N_S}}(s\mid\Phi_{\mathrm{BS}})
=
\prod_{\mathrm{BS}_i\in\Phi_{\mathrm{BS}}\setminus\{\mathrm{BS}_1\}}
\!\left[1-A_ip_b+A_ip_b\Psi_i(s)\right],
\end{equation}
\begin{equation}
\begin{aligned}
\Psi_i(s)
&=p_{\mathrm{LOS}}(r_i)
\left(1+\frac{sP_cG_m^2C_Lr_i^{-\alpha_L}}{m_L}\right)^{-m_L}\\
&\quad+[1-p_{\mathrm{LOS}}(r_i)]
\left(1+\frac{sP_cG_m^2C_Nr_i^{-\alpha_N}}{m_N}\right)^{-m_N}.
\end{aligned}
\end{equation}
}
\end{lemma}
\begin{IEEEproof}
Starting from \eqref{eq:ILS_INS_def}, conditioning on $\Phi_{\mathrm{BS}}$ fixes $r_i$ and the geometry-dependent indicator $A_i$, while $B_i$ remains random due to local beam scheduling. If $A_i=0$ or $B_i=0$, BS~$i$ contributes no direct interference. Otherwise, averaging over
the LoS/NLoS state and unit-mean Gamma fading using
$\mathbb{E}[e^{-ag_{\chi}}]=(1+a/m_\chi)^{-m_\chi}$
gives $\Psi_i(s)$. Averaging over
$B_i\sim\mathrm{Bernoulli}(p_b)$ yields the per-BS factor
$1-A_ip_b+A_ip_b\Psi_i(s)$. Independence of beam alignment,
propagation state, and fading across BSs then gives the product in
\eqref{eq:LT_direct_cond}.
\end{IEEEproof}

\begin{lemma}
\label{lem:target_LT}
Conditioned on $\Phi_{\mathrm{BS}}$, the LT of the target-reflected interference is
\par
{\small
\begin{equation}
\label{eq:LT_target_cond}
\mathcal{L}_{I_T}(s\mid\Phi_{\mathrm{BS}})
=
\prod_{\mathrm{BS}_i\in\Phi_{\mathrm{BS}},\,i>N_c}
\left[
1-p_{\mathrm{LOS}}(R_i)p_b
+\frac{p_{\mathrm{LOS}}(R_i)p_b}{1+s\beta_i}
\right],
\end{equation}
}
where
$\beta_i\triangleq
\frac{P_sG_m^2\lambda^2\sigma_{\mathrm{av,bi}}}
{(4\pi)^3R_1^{\alpha_L}R_i^{\alpha_L}}$.
\end{lemma}
\begin{IEEEproof}
Starting from \eqref{eq:IT_def}, conditioning on $\Phi_{\mathrm{BS}}$
fixes $R_1$ and $R_i$, and hence the corresponding coefficient
$K_i=P_sG_m^2\lambda^2/
[(4\pi)^3R_1^{\alpha_L}R_i^{\alpha_L}]$.
For each $i>N_c$, let
$q_i=p_{\mathrm{LOS}}(R_i)p_b$ denote the conditional probability that
BS~$i$ illuminates the target over a LoS path. If this event does not
occur, BS~$i$ contributes no target-reflected interference. Otherwise,
since $\sigma_{b,i}$ is exponential with mean
$\sigma_{\mathrm{av,bi}}$,
$\mathbb{E}[e^{-sK_i\sigma_{b,i}}]=(1+s\beta_i)^{-1}$, where
$\beta_i=K_i\sigma_{\mathrm{av,bi}}$. Thus, the per-BS conditional LT
is $1-q_i+q_i/(1+s\beta_i)$. Independence of illumination 
and RCS fluctuations across BSs then gives the product in
\eqref{eq:LT_target_cond}.
\end{IEEEproof}

Using Lemmas~\ref{lem:direct_LT} and~\ref{lem:target_LT} with the Gamma approximation, we obtain the conditional cooperative sensing success probability.
\begin{theorem}
\label{thm:cond_cov}
Conditioned on $\Phi_{\mathrm{BS}}$, the cooperative sensing success probability is
\par
{\small
\begin{equation}
\label{eq:cond_cov_result}
\begin{aligned}
\mathcal{P}_{\mathrm{S}}(\Phi_{\mathrm{BS}};\tau)
&\approx
\sum_{z=1}^{\infty}
\binom{k_{\mathrm{eff}}}{z}(-1)^{z+1}
\mathcal{L}_{I_T}(s_z\mid\Phi_{\mathrm{BS}})\\
&\quad\times
\mathcal{L}_{I_{L_S}+I_{N_S}}(s_z\mid\Phi_{\mathrm{BS}})
e^{-s_z(K_BTW+P_c\zeta_{\mathrm{SI}})},
\end{aligned}
\end{equation}
}
where $s_z\triangleq z\eta\tau/\theta_{\mathrm{eff}}$,
$\eta\triangleq[\Gamma(1+k_{\mathrm{eff}})]^{-1/k_{\mathrm{eff}}}$,
and $\binom{k_{\mathrm{eff}}}{z}$ denotes the generalized binomial
coefficient.
\end{theorem}

\begin{IEEEproof}
Let
$I_{\mathrm{tot}}\triangleq
I_T+I_{L_S}+I_{N_S}+K_BTW+P_c\zeta_{\mathrm{SI}}$.
For tractability, we treat $P_{\mathrm{tot}}$, $I_{L_S}+I_{N_S}$, and
$I_T$ as conditionally independent given $\Phi_{\mathrm{BS}}$.
Conditioned on $I_{\mathrm{tot}}$ and $\Phi_{\mathrm{BS}}$, the sensing
success event is $P_{\mathrm{tot}}>\tau I_{\mathrm{tot}}$. Using the
Gamma approximation in \eqref{eq:gamma_approx}, the Alzer-based
approximation \cite{alzer1997some} gives
$\bar F_{P_{\mathrm{tot}}\mid\Phi_{\mathrm{BS}}}(x)
\approx1-[1-\exp(-\eta x/\theta_{\mathrm{eff}})]^{k_{\mathrm{eff}}}$.
Setting $x=\tau I_{\mathrm{tot}}$ and applying the generalized binomial
expansion gives
$\sum_{z=1}^{\infty}\binom{k_{\mathrm{eff}}}{z}
(-1)^{z+1}e^{-s_zI_{\mathrm{tot}}}$.
Averaging over the remaining interference randomness conditioned on
$\Phi_{\mathrm{BS}}$, separating the deterministic noise and residual
SI terms, and applying Lemmas~\ref{lem:direct_LT}
and~\ref{lem:target_LT} yields \eqref{eq:cond_cov_result}.
\end{IEEEproof}
The conditional-independence approximation used in the proof is motivated by the fact that, for a neighboring BS, the direction toward the target generally differs from the direction toward the sensing BS, while useful cooperative echoes and out-of-cluster target-reflected interference arise from disjoint BS sets. Its accuracy is assessed through exact-geometry Monte Carlo simulations in Section~\ref{num_ress}.

\subsubsection{Cooperative Sensing Meta-Distribution}

Having obtained the conditional sensing success probability, we characterize its variability across targets. The cooperative sensing meta-distribution is the complementary cumulative distribution function (CCDF) of
$\mathcal{P}_{\mathrm{S}}(\Phi_{\mathrm{BS}};\tau)$
\cite{7345601,zhu2026meta}:
\par
{\small
\begin{equation}
\label{eq:meta_def}
\bar{F}_p(t)
\triangleq
\mathbb{P}_{\Phi_{\mathrm{BS}}}
\left(
\mathcal{P}_{\mathrm{S}}(\Phi_{\mathrm{BS}};\tau)> t
\right),
\qquad t\in[0,1].
\end{equation}
}
Thus, $\bar{F}_p(t)$ gives the fraction of targets whose conditional sensing success probability exceeds $t$.
Define the $b$-th moment of the conditional sensing success probability as
\begin{equation}
\label{eq:Mb_def}
M_b(\tau)
\triangleq
\mathbb{E}_{\Phi_{\mathrm{BS}}}
\left[
\mathcal{P}_{\mathrm{S}}(\Phi_{\mathrm{BS}};\tau)^b
\right],
\qquad b\in\mathbb{C}.
\end{equation}
Substituting \eqref{eq:cond_cov_result} into \eqref{eq:Mb_def} gives
\par
{\small
\begin{equation}
\label{eq:Mb_general}
\begin{aligned}
M_b(\tau)
&\approx
\mathbb{E}_{\Phi_{\mathrm{BS}}}
\Bigg[
\Bigg(
\sum_{z=1}^{\infty}
\binom{k_{\mathrm{eff}}}{z}(-1)^{z+1}
\mathcal{L}_{I_T}(s_z\mid\Phi_{\mathrm{BS}})\\
&\quad\times
\mathcal{L}_{I_{L_S}+I_{N_S}}(s_z\mid\Phi_{\mathrm{BS}})
e^{-s_z(K_BTW+P_c\zeta_{\mathrm{SI}})}
\Bigg)^b
\Bigg].
\end{aligned}
\end{equation}
}
The expectation in \eqref{eq:Mb_general} is only over the BS geometry
$\Phi_{\mathrm{BS}}$, since transmit-beam alignment, LoS/NLoS states,
fading, and RCS fluctuations are already averaged when computing
$\mathcal{P}_{\mathrm{S}}(\Phi_{\mathrm{BS}};\tau)$.
For $b=ju$, $u\in\mathbb{R}$,
$M_{ju}(\tau)=
\mathbb{E}_{\Phi_{\mathrm{BS}}}
[e^{ju\log\mathcal{P}_{\mathrm{S}}(\Phi_{\mathrm{BS}};\tau)}]$,
which is the characteristic function of
$\log\mathcal{P}_{\mathrm{S}}(\Phi_{\mathrm{BS}};\tau)$.
By the Gil--Pelaez inversion theorem \cite{7345601},
\par
{\small
\begin{equation}
\label{eq:meta_GP}
\bar{F}_p(t)
=
\frac{1}{2}
+
\frac{1}{\pi}
\int_{0}^{\infty}
\frac{\Im\!\left[
e^{-ju\log(t)}M_{ju}(\tau)
\right]}{u}\,du,
\qquad 0<t<1,
\end{equation}
}
where $\Im(\cdot)$ denotes the imaginary part and $j^2=-1$.

\subsubsection{Cooperative Sensing Coverage Probability and Ergodic Rate}

Besides the meta-distribution, we evaluate cooperative sensing coverage
$\mathbb{P}(\mathrm{SINR}_S>\tau)$ and sensing ergodic rate
$\mathbb{E}[\log(1+\mathrm{SINR}_S)]$, widely adopted system-level
ISAC metrics \cite{olson2023coverage,meng2024network,11270002},
as spatially averaged performance measures.

The first moment gives the spatially averaged sensing coverage probability
of the typical target,
$\bar{\mathcal{P}}_{\mathrm{S}}(\tau)
\triangleq\mathbb{P}(\mathrm{SINR}_S>\tau)=M_1(\tau)$.
Using the tail-integral identity for $\log(1+\mathrm{SINR}_S)$, the
aggregate sensing ergodic rate per BS is
\par
{\small
\begin{equation}
\label{eq:sensing_rate_M1}
\mathcal{R}_{\mathrm{S}}
=
N_b
\int_{0}^{\infty}
M_1(e^x-1)\,dx,
\end{equation}
}
where $M_1(e^x-1)$ denotes \eqref{eq:Mb_general} evaluated with
$b=1$ and $\tau=e^x-1$, and $N_b$ accounts for the $N_b$
simultaneous beams.

\subsection{Communication Analysis}
\label{ana_commm}

Consider a typical MU served by its nearest BS at distance $R_0$. The
serving beam is perfectly aligned with the MU, yielding gain $G_m$. To 
capture the angular gain variation of interfering beams, we adopt the cosine antenna
model \cite{yu2017coverage}
\begin{equation}
\label{ant_be_ga}
G(\theta)=
\begin{cases}
G_m \cos^2\!\left(\frac{\pi\theta}{2\theta_b}\right), & |\theta|\leq \theta_b,\\
0, & \text{otherwise}.
\end{cases}
\end{equation}
The downlink SINR is
\par
{\small
\begin{equation}
\label{eq:comm_sinr}
\mathrm{SINR}_C
=
\frac{P_cG_mC_Lg_{L,0}R_0^{-\alpha_L}}
{I_{L_C}+I_{N_C}+K_BTW},
\end{equation}
}
where $g_{L,0}$ is the desired LoS fading gain and, for
$\chi\in\{L,N\}$, $I_{\chi_C}$ is the aggregate interference from BS
beams whose links to the typical MU are in state $\chi$.

\begin{lemma}
\label{lem:comm_LT}
Conditioned on $R_0$, the LT of $I_{\chi_C}$ is
\par
{\small
\begin{equation}
\label{eq:LT_comm}
\mathcal{L}_{I_{\chi_C}}(s\mid R_0)
=
\exp\!\left[-\lambda_{\mathrm{BS}}N_b
\mathcal{J}_{\chi}(s,R_0)\right],
\quad \chi\in\{L,N\},
\end{equation}
}
\par
{\footnotesize
\begin{equation}
\label{eq:J_comm}
\mathcal{J}_{\chi}(s,R_0)
=
\int_{-\theta_b}^{\theta_b}\!\int_{R_0}^{\infty}
p_\chi(r)
\left[
1-
\left(
1+\frac{sP_cG(\theta)C_\chi}
{m_\chi r^{\alpha_\chi}}
\right)^{-m_\chi}
\right]r\,dr\,d\theta,
\end{equation}
}
\noindent
where $p_L(r)=p_{\mathrm{LOS}}(r)$ and
$p_N(r)=1-p_{\mathrm{LOS}}(r)$.
\end{lemma}
\begin{IEEEproof}
Conditioning on $R_0$ excludes interfering BSs within radius $R_0$.
As in Lemma~\ref{lem:direct_LT}, averaging the unit-mean Gamma fading
gives $(1+a/m_\chi)^{-m_\chi}$. Averaging over the cosine-pattern
support in \eqref{ant_be_ga} and applying the PPP probability generating functional (PGFL) for $r\ge R_0$,
accounting for the $N_b$ beams per BS, yields
\eqref{eq:LT_comm}.
\end{IEEEproof}

\begin{theorem}
\label{thm:comm_cov}
For communication SINR threshold $\tau_c$, the coverage probability
$\mathcal{P}_C(\tau_c)\triangleq
\mathbb{P}(\mathrm{SINR}_C>\tau_c)$ is
\par
{\small
\begin{equation}
\label{eq:comm_cov}
\begin{aligned}
\mathcal{P}_C(\tau_c)
&\approx
\int_0^\infty
\sum_{z=1}^{m_L}
(-1)^{z+1}\binom{m_L}{z}
e^{-s_{c,z}(r)K_BTW}\\
&\quad\times
\mathcal{L}_{I_{L_C}}(s_{c,z}(r)\mid r)
\mathcal{L}_{I_{N_C}}(s_{c,z}(r)\mid r)
f_{R_0}(r)\,dr ,
\end{aligned}
\end{equation}
}
where $m_L$ is a positive integer,
$s_{c,z}(r)\triangleq
zk_L\tau_cr^{\alpha_L}/(P_cG_mC_L)$,
$k_L\triangleq m_L(m_L!)^{-1/m_L}$, and
$f_{R_0}(r)=2\pi\lambda_{\mathrm{BS}}r
e^{-\pi\lambda_{\mathrm{BS}}r^2}$.
\end{theorem}

\begin{IEEEproof}
Conditioned on $R_0=r$, applying the Alzer-based approximation
\cite{alzer1997some} to the desired LoS fading gain and using
Lemma~\ref{lem:comm_LT} gives the conditional coverage probability.
Averaging over $R_0$ yields \eqref{eq:comm_cov}.
\end{IEEEproof}
Using the tail-integral identity, the aggregate communication ergodic
rate per BS is
\par
{\small
\begin{equation}
\label{eq:comm_rate}
\mathcal{R}_C
=
\frac{N_b(T_t-T_s)}{T_t}
\int_0^\infty
\mathcal{P}_C(e^x-1)\,dx ,
\end{equation}
}
where $(T_t-T_s)/T_t$ is the communication duty factor and $N_b$
the number of simultaneously served MUs.

\section{Numerical Results}
\label{num_ress}

We validate the analysis using exact-geometry Monte Carlo simulations, where echo powers, interference, and cooperative links are evaluated in each realization without the analytical approximations used in the derivations. Results are based on $2\times10^5$ independent realizations over a $25$~km$^2$ area at 28~GHz. Following \cite{olson2023coverage}, we set $R_{\max}=3R_{\rm eff}$ with $R_{\rm eff}=\sqrt{1/(\pi\lambda_{\mathrm{BS}})}$. The mmWave parameters follow established studies \cite{yu2017coverage,rebato2019stochastic}. Unless stated otherwise, we use $W=200$~MHz, $N_b=8$, $\theta_b=18^\circ$, $G_m=20$~dBi, $\alpha_L=2$, $\alpha_N=4$, $C_L=-61.4$~dB, $C_N=-72$~dB, $m_L=3$, $m_N=2$, $\sigma_{\mathrm{av,mono}}=1$~m$^2$, $\sigma_{\mathrm{av,bi}}=0.7$~m$^2$, $\eta_b=0.0149$~m$^{-1}$, $\lambda_{\mathrm{BS}}=70$~BS/km$^2$, $\gamma=0.7$, $T=300$~K, and per-beam slot-averaged transmit power $E/T_t=1$~W. We set $\zeta_{\mathrm{SI}}=10^{-10}$, corresponding to $100$~dB SI suppression, consistent with experimentally demonstrated suppression levels summarized in \cite{11145123}.

\begin{figure*}[!t]
\centering
\subfloat[\label{sen_sim}]{\input{sen_cov}}\hfill
\subfloat[\label{meta_dis}]{\input{meta}}\hfill
\subfloat[\label{bs_sen}]{\input{tot_rate}}\hfill
\subfloat[\label{pw_com}]{\input{pow_comp}}
\caption{(a) Spatially averaged sensing coverage probability versus sensing SINR threshold. (b) Sensing meta-distribution at a sensing SINR threshold of $5$~dB. (c) Aggregate sensing and communication ergodic rates per BS versus BS density. (d) Aggregate sensing and communication ergodic rates per BS versus the energy-splitting factor $\gamma$.}
\label{fig:four_subfigs}
\end{figure*}

Fig.~\ref{sen_sim} plots the average sensing coverage probability versus the SINR threshold. Close analytical--simulation agreement supports the Bernoulli-gated cooperation model and the accuracy of the moment-matched Gamma and conditional-independence approximations in Theorem~\ref{thm:cond_cov}. The $N_c=1$ case provides the monostatic baseline, while increasing $N_c$ adds bistatic-echo opportunities and excludes more nearby BSs from target-reflected interference. Gains diminish beyond $N_c=4$ because additional candidates are farther, with higher path loss and lower LoS probability. Hence, the results suggest selecting the smallest $N_c$ that meets the desired sensing performance, since further increasing the cluster size raises echo-processing burden while providing limited gain.

Fig.~\ref{meta_dis} shows the sensing meta-distribution, with close analytical--simulation agreement validating the meta-distribution characterization. Cooperation markedly increases the fraction of targets whose conditional sensing success probability exceeds $t$, particularly in the high-reliability regime. At $t=0.9$, increasing $N_c$ from $1$ to $6$ raises this fraction from approximately $20\%$ to $77\%$. Thus, opportunistic cooperation strengthens target-level reliability beyond what average coverage alone reveals, substantially increasing the fraction of targets meeting stringent sensing requirements relevant to safety-critical 6G applications.

Fig.~\ref{bs_sen} plots the average rates versus BS density. Opportunistic cooperation improves the sensing rate across the considered densities. At low $\lambda_{\mathrm{BS}}$, densification benefits sensing by shortening sensing paths and increasing the LoS availability of candidate links. Because sensing echoes traverse both illumination and return paths, shortening these paths can yield a stronger received-power benefit. At higher densities, rising LoS interference increasingly erodes these benefits, leading to sensing-rate saturation. In contrast, the communication rate reaches its optimum at a lower density and then decreases as the network becomes interference-limited. These different density dependencies highlight the need to balance both functions when selecting $\lambda_{\mathrm{BS}}$.

 Fig.~\ref{pw_com} examines the energy-splitting factor $\gamma$. Increasing $\gamma$ strengthens sensing echoes while reducing residual SI through lower communication power, thereby improving the sensing rate. Opportunistic cooperation improves sensing, as shown by the separation between the $N_c=1$ and $N_c=6$ curves. In contrast, the communication rate varies slowly over $\gamma$ because, in the interference-limited regime, decreasing communication power scales the desired signal and aggregate interference similarly. As $\gamma$ approaches one, communication power becomes too small and thermal noise becomes non-negligible, causing the rate to drop sharply. Thus, an appropriately sensing-biased energy allocation can provide substantial sensing gains with limited communication-rate loss.

\section{Conclusion}
\label{con_pp}

This paper proposed an opportunistic cooperative sensing architecture for mmWave ISAC networks that harvests bistatic echoes from already-active neighboring-BS transmissions without dedicated air-interface resources for cooperation. The required coordination can leverage existing 5G transport and MEC infrastructure for synchronization and local echo processing. A stochastic-geometry framework characterized conditional sensing reliability, its meta-distribution, and average sensing and communication performance. The meta-distribution revealed strong target-level gains, especially in the high-reliability regime. Incremental gains diminish as cluster size increases, motivating selection of the smallest cluster that meets the sensing requirement. Results further revealed distinct sensing--communication tradeoffs: sensing benefits from additional densification beyond the communication-optimal density, while an appropriately sensing-biased energy allocation provides substantial sensing gains with limited communication-rate loss. These findings demonstrate the potential of opportunistically exploiting existing multi-BS transmissions to enhance reliable sensing in future 6G networks.

\bibliographystyle{IEEEtran}
\bibliography{bibliography.bib}

\end{document}

%% file: sen_cov.tex
%
%
\definecolor{mycolor1}{rgb}{0.00000,0.44700,0.74100}%
\definecolor{mycolor2}{rgb}{0.85000,0.32500,0.09800}%
\definecolor{mycolor3}{rgb}{0.92900,0.69400,0.12500}%
\definecolor{mycolor4}{rgb}{0.49400,0.18400,0.55600}%

\begin{tikzpicture}[scale=0.3145, transform shape,font=\LARGE]

\begin{axis}[%
width=4.521in,
height=3.566in,
at={(0.758in,0.481in)},
scale only axis,
xmin=-10,
xmax=10,
xlabel style={font=\huge\color{white!15!black}},
xlabel={SINR Threshold (dB)},
ymin=0.65,
ymax=1,
ylabel style={font=\huge\color{white!15!black}},
ylabel={Sensing Coverage Probability},
axis background/.style={fill=white},
xmajorgrids,
ymajorgrids,
legend style={font=\huge,at={(0.02,0.02)}, anchor=south west, legend cell align=left, align=left, draw=white!15!black},
  grid=both,
]
\addplot [line width=2.0pt, color=mycolor1]
  table[row sep=crcr]{%
-10	0.983086159178734\\
-9	0.979677168067793\\
-8	0.975633131829769\\
-7	0.97083972276571\\
-6	0.965165665202397\\
-5	0.958464213809493\\
-4	0.950576916091612\\
-3	0.94134005680099\\
-2	0.930593818235392\\
-1	0.918193614755333\\
0	0.904022363138235\\
1	0.888001842581752\\
2	0.870101055424644\\
3	0.850339836374964\\
4	0.828786880598728\\
5	0.805552607627072\\
6	0.780778418241105\\
7	0.754624553429649\\
8	0.727258785442464\\
9	0.698847689724322\\
10	0.669551524374958\\
};
\addlegendentry{$N_c$=1}

\addplot [line width=2.0pt, color=mycolor2]
  table[row sep=crcr]{%
-10	0.988506870406477\\
-9	0.986263964413772\\
-8	0.983623234731168\\
-7	0.980509110035264\\
-6	0.976829847629583\\
-5	0.972476104875753\\
-4	0.967320657417105\\
-3	0.961219749757977\\
-2	0.954016510145181\\
-1	0.945546674167915\\
0	0.935646519750562\\
1	0.924162456156801\\
2	0.91096123531626\\
3	0.895939419840419\\
4	0.879030698819674\\
5	0.860209960640783\\
6	0.839493651723899\\
7	0.816936696142552\\
8	0.792626914115763\\
9	0.766678303306149\\
10	0.739224664425108\\
};
\addlegendentry{$N_c$=2}

\addplot [line width=2.0pt, color=mycolor3]
  table[row sep=crcr]{%
-10	0.99181608392777\\
-9	0.990338649598127\\
-8	0.988641370541751\\
-7	0.986686103236656\\
-6	0.98442317071092\\
-5	0.98178886303358\\
-4	0.978703177787802\\
-3	0.975068073803682\\
-2	0.970766563810961\\
-1	0.965662997120027\\
0	0.959604859283948\\
1	0.952426324051041\\
2	0.94395362334567\\
3	0.934012058218946\\
4	0.922434188666567\\
5	0.909068474948015\\
6	0.893787480221133\\
7	0.876494761729303\\
8	0.857129817201999\\
9	0.835670891663103\\
10	0.812135973746484\\
};
\addlegendentry{$N_c$=4}

\addplot [line width=2.0pt, color=mycolor4]
  table[row sep=crcr]{%
-10	0.99285940183078\\
-9	0.991635895508474\\
-8	0.990257545302997\\
-7	0.988702096140728\\
-6	0.986938684882787\\
-5	0.984925409395843\\
-4	0.982606885419395\\
-3	0.979911934086136\\
-2	0.976751582706578\\
-1	0.973017604030086\\
0	0.968581858873526\\
1	0.963296736923061\\
2	0.956996994488249\\
3	0.949503238584272\\
4	0.940627176302443\\
5	0.930178530009322\\
6	0.917973245745836\\
7	0.903842372508411\\
8	0.887640867358232\\
9	0.869255667202925\\
10	0.848612654114408\\
};
\addlegendentry{$N_c$=6}

\addplot [color=white, draw=none, mark=x, thick, mark size=6pt, mark options={ black}]
  table[row sep=crcr]{%
-10	0.983312\\
-9	0.979762\\
-8	0.975868\\
-7	0.971152\\
-6	0.965742\\
-5	0.959132\\
-4	0.9512\\
-3	0.94211\\
-2	0.931608\\
-1	0.91928\\
0	0.905404\\
1	0.889128\\
2	0.87135\\
3	0.851844\\
4	0.830674\\
5	0.8078\\
6	0.783262\\
7	0.757664\\
8	0.730216\\
9	0.701914\\
10	0.67322\\
};
\addlegendentry{Simulations}

\addplot [color=white, draw=none, mark=x, thick, mark size=6pt, mark options={ black}]
  table[row sep=crcr]{%
-10	0.98826\\
-9	0.985774\\
-8	0.983214\\
-7	0.97998\\
-6	0.976394\\
-5	0.972024\\
-4	0.966746\\
-3	0.960674\\
-2	0.953494\\
-1	0.945038\\
0	0.935324\\
1	0.923708\\
2	0.910572\\
3	0.895722\\
4	0.879192\\
5	0.860744\\
6	0.840288\\
7	0.818102\\
8	0.793808\\
9	0.7679\\
10	0.741288\\
};

\addplot [color=white, draw=none, mark=x, thick, mark size=6pt, mark options={ black}]
  table[row sep=crcr]{%
-10	0.991572\\
-9	0.98985\\
-8	0.988168\\
-7	0.986098\\
-6	0.98386\\
-5	0.981142\\
-4	0.977922\\
-3	0.974168\\
-2	0.969794\\
-1	0.964698\\
0	0.958832\\
1	0.95131\\
2	0.942702\\
3	0.93291\\
4	0.9217\\
5	0.908502\\
6	0.89333\\
7	0.876378\\
8	0.857092\\
9	0.835914\\
10	0.813006\\
};

\addplot [color=white, draw=none, mark=x, thick, mark size=6pt, mark options={ black}]
  table[row sep=crcr]{%
-10	0.992712\\
-9	0.991268\\
-8	0.989908\\
-7	0.988178\\
-6	0.986416\\
-5	0.984282\\
-4	0.981868\\
-3	0.978998\\
-2	0.975712\\
-1	0.97193\\
0	0.967504\\
1	0.96188\\
2	0.955444\\
3	0.948\\
4	0.939272\\
5	0.928894\\
6	0.916882\\
7	0.903102\\
8	0.886942\\
9	0.868784\\
10	0.84876\\
};

\end{axis}
\end{tikzpicture}%

%% file: tot_rate.tex
%
%
\definecolor{mycolor1}{rgb}{0.00000,0.44700,0.74100}%
\definecolor{mycolor2}{rgb}{0.85000,0.32500,0.09800}%
\definecolor{mycolor3}{rgb}{0.92900,0.69400,0.12500}%
\definecolor{mycolor4}{rgb}{0.49400,0.18400,0.55600}%
\definecolor{mycolor5}{rgb}{0.10,0.50,0.25}%
\begin{tikzpicture}[scale=0.316, transform shape,font=\LARGE]

\begin{axis}[%
width=4.521in,
height=3.559in,
at={(0.758in,0.488in)},
scale only axis,
xmin=0.5,
xmax=100,
xtick={10,20,30,40,50,60,70,80,90,100},
xlabel style={font=\huge\color{white!15!black}},
xlabel={$\text{BS density (BS/km}^2\text{)}$},
ymin=7,
ymax=55,
ylabel style={font=\huge\color{white!15!black}},
ylabel={Average Rate [nats/s/Hz]},
axis background/.style={fill=white},
xmajorgrids,
ymajorgrids,
legend style={font=\huge,at={(0.75,0.5)},legend cell align=left, align=left, draw=white!15!black}
]

\addplot [color=mycolor1, line width=2.0pt,  mark options={solid, mycolor1}]
  table[row sep=crcr]{%
0.1	7.66589164591497\\
2.13877551020408	12.2028000045528\\
4.17755102040816	17.2221548985749\\
6.21632653061225	20.504348041001\\
8.25510204081633	22.5930538351939\\
10.2938775510204	24.413964288374\\
12.3326530612245	25.9670794005415\\
14.3714285714286	27.201253622161\\
16.4102040816327	28.2580074328544\\
18.4489795918367	28.9724652243295\\
20.4877551020408	29.6522957704515\\
22.5265306122449	30.2041047424825\\
24.565306122449	30.6565891065076\\
26.6040816326531	31.0651796991921\\
28.6428571428571	31.4088805002552\\
30.6816326530612	31.8063164780107\\
32.7204081632653	32.1461128043411\\
34.7591836734694	32.3350104748087\\
36.7979591836735	32.3896063443771\\
38.8367346938776	32.3900186895899\\
40.8755102040816	32.5145439559106\\
42.9142857142857	32.7172805559134\\
44.9530612244898	32.7634910391629\\
46.9918367346939	32.7473011281273\\
49.030612244898	32.69670616489\\
51.069387755102	32.7260694298873\\
53.1081632653061	32.8271282504487\\
55.1469387755102	32.802497480538\\
57.1857142857143	32.674678105707\\
59.2244897959184	32.5956382360797\\
61.2632653061224	32.513883138123\\
63.3020408163265	32.4025802119774\\
65.3408163265306	32.2905280819994\\
67.3795918367347	32.1578139608592\\
69.4183673469388	32.0357629455202\\
71.4571428571429	31.9418094655549\\
73.4959183673469	31.8014419014384\\
75.534693877551	31.7038299055029\\
77.5734693877551	31.5998124251528\\
79.6122448979592	31.502919066749\\
81.6510204081633	31.4659858398813\\
83.6897959183673	31.2574865219389\\
85.7285714285714	31.0061297616614\\
87.7673469387755	30.8345455573491\\
89.8061224489796	30.6569843285409\\
91.8448979591837	30.5314215973634\\
93.8836734693878	30.4586106338192\\
95.9224489795918	30.3263744905813\\
97.9612244897959	30.1545714377808\\
100	29.9432014754177\\
};
\addlegendentry{$N_c$=1 Sens.}

\addplot [color=mycolor2, line width=2.0pt,  mark options={solid, mycolor2}]
  table[row sep=crcr]{%
0.1	7.67235888672785\\
2.13877551020408	12.2712704305853\\
4.17755102040816	17.4174556832979\\
6.21632653061225	20.8478756880484\\
8.25510204081633	23.1124492860746\\
10.2938775510204	25.0997484595076\\
12.3326530612245	26.8097732083474\\
14.3714285714286	28.1908322154038\\
16.4102040816327	29.389372100358\\
18.4489795918367	30.2411415219373\\
20.4877551020408	31.0787306490434\\
22.5265306122449	31.7852565590819\\
24.565306122449	32.3780403294896\\
26.6040816326531	32.9133112983869\\
28.6428571428571	33.3514122251134\\
30.6816326530612	33.8515299265543\\
32.7204081632653	34.3027684866929\\
34.7591836734694	34.5929832867091\\
36.7979591836735	34.7275693661625\\
38.8367346938776	34.8325966472282\\
40.8755102040816	35.043592753286\\
42.9142857142857	35.3404321311874\\
44.9530612244898	35.4667175527525\\
46.9918367346939	35.5183914931791\\
49.030612244898	35.515104174636\\
51.069387755102	35.6302436576844\\
53.1081632653061	35.8020096491446\\
55.1469387755102	35.8342275699986\\
57.1857142857143	35.7713768999417\\
59.2244897959184	35.7531488518161\\
61.2632653061224	35.7162467309601\\
63.3020408163265	35.6544784385789\\
65.3408163265306	35.5676946751293\\
67.3795918367347	35.4532785821633\\
69.4183673469388	35.3642605100978\\
71.4571428571429	35.3243095278992\\
73.4959183673469	35.2332698547067\\
75.534693877551	35.1794726624977\\
77.5734693877551	35.0990494574183\\
79.6122448979592	35.0206956105234\\
81.6510204081633	35.0082814756942\\
83.6897959183673	34.8165899877872\\
85.7285714285714	34.5733011432202\\
87.7673469387755	34.4117082148623\\
89.8061224489796	34.2391198468553\\
91.8448979591837	34.1461308661363\\
93.8836734693878	34.1114602488478\\
95.9224489795918	34.0021275208043\\
97.9612244897959	33.8453480613882\\
100	33.6411218705994\\
};
\addlegendentry{$N_c$=2 Sens.}

\addplot [color=mycolor3, line width=2.0pt,  mark options={solid, mycolor3}]
  table[row sep=crcr]{%
0.1	7.67246990711577\\
2.13877551020408	12.2811991501957\\
4.17755102040816	17.4699470151274\\
6.21632653061225	20.967418872541\\
8.25510204081633	23.3405972917302\\
10.2938775510204	25.4416978748569\\
12.3326530612245	27.2707206219212\\
14.3714285714286	28.7760515307321\\
16.4102040816327	30.1108068982524\\
18.4489795918367	31.0901220778986\\
20.4877551020408	32.0822761505342\\
22.5265306122449	32.9424245860372\\
24.565306122449	33.6890295367363\\
26.6040816326531	34.3663821986778\\
28.6428571428571	34.9354098226247\\
30.6816326530612	35.5661092453121\\
32.7204081632653	36.1648418181376\\
34.7591836734694	36.5969846967544\\
36.7979591836735	36.8603668023971\\
38.8367346938776	37.0944367199988\\
40.8755102040816	37.4345959539442\\
42.9142857142857	37.8605643601635\\
44.9530612244898	38.0961579758563\\
46.9918367346939	38.2640838194349\\
49.030612244898	38.3612026894216\\
51.069387755102	38.5999966884704\\
53.1081632653061	38.8911648636751\\
55.1469387755102	39.021761015403\\
57.1857142857143	39.0539418416855\\
59.2244897959184	39.1366979897413\\
61.2632653061224	39.1921468702923\\
63.3020408163265	39.218750649474\\
65.3408163265306	39.1960142660422\\
67.3795918367347	39.1349787231909\\
69.4183673469388	39.1077619892013\\
71.4571428571429	39.1507871580297\\
73.4959183673469	39.1415825236209\\
75.534693877551	39.1720084250403\\
77.5734693877551	39.1524638000904\\
79.6122448979592	39.1402289055935\\
81.6510204081633	39.1787741185425\\
83.6897959183673	39.0364742764699\\
85.7285714285714	38.8272655221664\\
87.7673469387755	38.69919254925\\
89.8061224489796	38.5596391852078\\
91.8448979591837	38.5254281126068\\
93.8836734693878	38.5432932688895\\
95.9224489795918	38.4811848569216\\
97.9612244897959	38.3677310120944\\
100	38.2029317344076\\
};
\addlegendentry{$N_c$=4 Sens.}

\addplot [color=mycolor4, line width=2.0pt,  mark options={solid, rotate=180, mycolor4}]
  table[row sep=crcr]{%
0.1	7.67247027744619\\
2.13877551020408	12.281664752646\\
4.17755102040816	17.4749651193763\\
6.21632653061225	20.9832245454929\\
8.25510204081633	23.3803850378392\\
10.2938775510204	25.5124143663584\\
12.3326530612245	27.3793125310505\\
14.3714285714286	28.9301961896993\\
16.4102040816327	30.3179018330944\\
18.4489795918367	31.351570545049\\
20.4877551020408	32.410521697424\\
22.5265306122449	33.3394244973055\\
24.565306122449	34.1572424484246\\
26.6040816326531	34.9073039404117\\
28.6428571428571	35.5505993772058\\
30.6816326530612	36.2554446401969\\
32.7204081632653	36.9371693893944\\
34.7591836734694	37.4505982794708\\
36.7979591836735	37.7877953255886\\
38.8367346938776	38.1012105536986\\
40.8755102040816	38.528138006179\\
42.9142857142857	39.0390449541555\\
44.9530612244898	39.3511852772739\\
46.9918367346939	39.5949130734167\\
49.030612244898	39.76180646233\\
51.069387755102	40.0785243011063\\
53.1081632653061	40.4529812827077\\
55.1469387755102	40.6626744822301\\
57.1857142857143	40.7666267481593\\
59.2244897959184	40.9184822390578\\
61.2632653061224	41.0451237716393\\
63.3020408163265	41.1386051003473\\
65.3408163265306	41.1805668252929\\
67.3795918367347	41.180723089209\\
69.4183673469388	41.2080118603977\\
71.4571428571429	41.3156395049442\\
73.4959183673469	41.3698990827125\\
75.534693877551	41.4665808529425\\
77.5734693877551	41.5037875251579\\
79.6122448979592	41.5451122696399\\
81.6510204081633	41.6293480599869\\
83.6897959183673	41.5347147961587\\
85.7285714285714	41.3712245628125\\
87.7673469387755	41.2883646779066\\
89.8061224489796	41.1907852348458\\
91.8448979591837	41.2038181139362\\
93.8836734693878	41.2700609676571\\
95.9224489795918	41.2545195691102\\
97.9612244897959	41.1866555849558\\
100	41.0664690151938\\
};
\addlegendentry{$N_c$=6 Sens.}

\addplot [color=mycolor5, line width=2.0pt, mark options={solid, blue}]
  table[row sep=crcr]{%
0.1	18.2544378492737\\
2.5469387755102	46.2840551420033\\
4.99387755102041	50.8462194126251\\
7.44081632653061	52.9724635784221\\
9.88775510204082	54.0211344209977\\
12.334693877551	54.4554932863446\\
14.7816326530612	54.4964882327262\\
17.2285714285714	54.2690536688719\\
19.6755102040816	53.8515281002294\\
22.1224489795918	53.2965309121211\\
24.569387755102	52.6411590409575\\
27.0163265306122	51.9125062878427\\
29.4632653061225	51.1308899833093\\
31.9102040816327	50.3118536099187\\
34.3571428571429	49.4674706950415\\
36.8040816326531	48.6072276811539\\
39.2510204081633	47.7386413798681\\
41.6979591836735	46.8677028592426\\
44.1448979591837	45.9992040725607\\
46.5918367346939	45.1369830124038\\
49.0387755102041	44.2841109633438\\
51.4857142857143	43.4430377159486\\
53.9326530612245	42.6157056310832\\
56.3795918367347	41.8036403107443\\
58.8265306122449	41.0080234039294\\
61.2734693877551	40.2297515919442\\
63.7204081632653	39.4694847737749\\
66.1673469387755	38.7276856351514\\
68.6142857142857	38.0046525047319\\
71.0612244897959	37.3005465902748\\
73.5081632653061	36.6154147648306\\
75.9551020408163	35.9492087955525\\
78.4020408163265	35.3018014063442\\
80.8489795918367	34.6729999433945\\
83.2959183673469	34.0625579718638\\
85.7428571428571	33.4701850715105\\
88.1897959183674	32.895555153701\\
90.6367346938775	32.3383136813576\\
93.0836734693877	31.7980836820364\\
95.530612244898	31.2744709357504\\
97.9775510204081	30.7670684369421\\
100.424489795918	30.2754601079738\\
102.871428571429	29.7992240955214\\
105.318367346939	29.337935448924\\
107.765306122449	28.8911684614649\\
110.212244897959	28.4584986570965\\
112.659183673469	28.0395044543018\\
115.10612244898	27.6337685349666\\
117.55306122449	27.2408790599348\\
120	26.8604305839274\\
};
\addlegendentry{Comm.}

\addplot [color=white, draw=none, mark=x, thick, mark size=6pt,  mark repeat=3, mark options={ black}]
  table[row sep=crcr]{%
0.1	21.3061739393163\\
2.5469387755102	45.6721321114541\\
4.99387755102041	50.1983210742652\\
7.44081632653061	52.3495663261018\\
9.88775510204082	53.4187509897409\\
12.334693877551	53.8218500619992\\
14.7816326530612	53.8658538345505\\
17.2285714285714	53.6476764408646\\
19.6755102040816	53.244820622203\\
22.1224489795918	52.6647314471185\\
24.569387755102	52.0126611816931\\
27.0163265306122	51.2659742973668\\
29.4632653061225	50.5365300498766\\
31.9102040816327	49.7684918624361\\
34.3571428571429	48.8777749925812\\
36.8040816326531	47.9999234678147\\
39.2510204081633	47.111727852895\\
41.6979591836735	46.2357351727462\\
44.1448979591837	45.4211066354877\\
46.5918367346939	44.603004169709\\
49.0387755102041	43.7612271086697\\
51.4857142857143	42.8318706405374\\
53.9326530612245	41.9700005420379\\
56.3795918367347	41.2655015409037\\
58.8265306122449	40.4449348823949\\
61.2734693877551	39.6929448225925\\
63.7204081632653	38.9040297687286\\
66.1673469387755	38.1409910259298\\
68.6142857142857	37.3571166306249\\
71.0612244897959	36.7499123365943\\
73.5081632653061	36.0521334504304\\
75.9551020408163	35.4630993687023\\
78.4020408163265	34.7502230615575\\
80.8489795918367	34.0702697253769\\
83.2959183673469	33.5124632993799\\
85.7428571428571	32.9279322938396\\
88.1897959183674	32.3239569816843\\
90.6367346938775	31.7740085294804\\
93.0836734693877	31.211806882769\\
95.530612244898	30.695932466719\\
97.9775510204081	30.1708695599535\\
100.424489795918	29.7761220550324\\
102.871428571429	29.2414792725401\\
105.318367346939	28.7277034687487\\
107.765306122449	28.3399938056233\\
110.212244897959	27.875278647122\\
112.659183673469	27.4894546889474\\
115.10612244898	27.0851054655471\\
117.55306122449	26.7336604461295\\
120	26.279388184796\\
};
\addlegendentry{Simulations}

\addplot [color=white, draw=none, mark=x, thick, mark size=6pt,    mark repeat=4, mark options={ black}]
  table[row sep=crcr]{%
0.1	15.741846395689\\
2.13877551020408	12.7498683637159\\
4.17755102040816	17.0881872221716\\
6.21632653061225	20.254391471784\\
8.25510204081633	22.4319088360115\\
10.2938775510204	24.4407093643744\\
12.3326530612245	25.7766375302616\\
14.3714285714286	27.1135199329402\\
16.4102040816327	27.9757255649471\\
18.4489795918367	28.7108778800774\\
20.4877551020408	29.6387068482842\\
22.5265306122449	30.2926359448331\\
24.565306122449	30.7040534925263\\
26.6040816326531	31.1836723512428\\
28.6428571428571	31.4776866900313\\
30.6816326530612	31.7880908064994\\
32.7204081632653	32.0496133236246\\
34.7591836734694	32.2327893431269\\
36.7979591836735	32.334815493278\\
38.8367346938776	32.601809093581\\
40.8755102040816	32.6159357847973\\
42.9142857142857	32.7803090952578\\
44.9530612244898	32.8625829267003\\
46.9918367346939	32.7400022823154\\
49.030612244898	32.7671435251572\\
51.069387755102	32.9586695858846\\
53.1081632653061	32.9799049995691\\
55.1469387755102	32.6016734655553\\
57.1857142857143	32.6797063687639\\
59.2244897959184	32.513339907517\\
61.2632653061224	32.4329242298217\\
63.3020408163265	32.3902935620505\\
65.3408163265306	32.4436520631155\\
67.3795918367347	32.4219940261049\\
69.4183673469388	32.1427198765602\\
71.4571428571429	32.1788951389267\\
73.4959183673469	31.8939260230136\\
75.534693877551	31.9381232561646\\
77.5734693877551	31.5614645782422\\
79.6122448979592	31.454379312941\\
81.6510204081633	31.5755477456871\\
83.6897959183673	31.2989488118966\\
85.7285714285714	31.2703448205009\\
87.7673469387755	30.924044671279\\
89.8061224489796	30.692958344787\\
91.8448979591837	30.6173545792633\\
93.8836734693878	30.4930197447831\\
95.9224489795918	30.5299941685504\\
97.9612244897959	30.1490688214474\\
100	29.9614568576006\\
};

  \addplot [color=white, draw=none, mark=x, thick, mark size=6pt,    mark repeat=4,mark options={ black}]
  table[row sep=crcr]{%
0.1	15.7789362783011\\
2.13877551020408	12.8137647559559\\
4.17755102040816	17.2751680475038\\
6.21632653061225	20.6200131351019\\
8.25510204081633	22.9019766579409\\
10.2938775510204	25.1106417526622\\
12.3326530612245	26.6024102452157\\
14.3714285714286	28.0925249552443\\
16.4102040816327	29.0984626249188\\
18.4489795918367	29.9917077608446\\
20.4877551020408	31.0218790659537\\
22.5265306122449	31.7589640907739\\
24.565306122449	32.357910093407\\
26.6040816326531	33.0068630373895\\
28.6428571428571	33.3911314708262\\
30.6816326530612	33.7586431996733\\
32.7204081632653	34.1487458824773\\
34.7591836734694	34.4767117626878\\
36.7979591836735	34.7123124638158\\
38.8367346938776	34.9785980132582\\
40.8755102040816	35.0864118135991\\
42.9142857142857	35.4721226729027\\
44.9530612244898	35.5575394541032\\
46.9918367346939	35.4594958368225\\
49.030612244898	35.5422105555446\\
51.069387755102	35.7613846649193\\
53.1081632653061	35.9297319545856\\
55.1469387755102	35.6293044795953\\
57.1857142857143	35.6920012499951\\
59.2244897959184	35.6991288580126\\
61.2632653061224	35.5512618471358\\
63.3020408163265	35.609483848583\\
65.3408163265306	35.7042653497393\\
67.3795918367347	35.6377795917702\\
69.4183673469388	35.4016577370996\\
71.4571428571429	35.5396540430531\\
73.4959183673469	35.2807605754456\\
75.534693877551	35.3583437319004\\
77.5734693877551	34.9498209869048\\
79.6122448979592	34.8855506063055\\
81.6510204081633	35.1258609214118\\
83.6897959183673	34.81969010014\\
85.7285714285714	34.7475047629264\\
87.7673469387755	34.4083369540306\\
89.8061224489796	34.3313621446574\\
91.8448979591837	34.2845508236859\\
93.8836734693878	34.1311557498749\\
95.9224489795918	34.0804210748914\\
97.9612244897959	33.802360043099\\
100	33.6793542561906\\
};

  \addplot [color=white, draw=none, mark=x, thick, mark size=6pt,    mark repeat=4,mark options={ black}]
  table[row sep=crcr]{%
0.1	15.7819325857359\\
2.13877551020408	12.8227794694421\\
4.17755102040816	17.3248675469547\\
6.21632653061225	20.7380856716787\\
8.25510204081633	23.1174030786158\\
10.2938775510204	25.4359740639786\\
12.3326530612245	27.0499622616314\\
14.3714285714286	28.6553621086693\\
16.4102040816327	29.8525356256677\\
18.4489795918367	30.8526647495287\\
20.4877551020408	32.0358216381709\\
22.5265306122449	32.9384130386856\\
24.565306122449	33.6340440235778\\
26.6040816326531	34.4201292461894\\
28.6428571428571	35.0723535507698\\
30.6816326530612	35.529355559902\\
32.7204081632653	35.9551504241425\\
34.7591836734694	36.4883224748738\\
36.7979591836735	36.8353369548846\\
38.8367346938776	37.213303384017\\
40.8755102040816	37.419258937325\\
42.9142857142857	37.9561541939809\\
44.9530612244898	38.2159177118051\\
46.9918367346939	38.2666853291999\\
49.030612244898	38.4125270319108\\
51.069387755102	38.7387629019988\\
53.1081632653061	39.0214752162317\\
55.1469387755102	38.793854431646\\
57.1857142857143	39.0670426201511\\
59.2244897959184	39.0833860194202\\
61.2632653061224	39.0042446156476\\
63.3020408163265	39.1341716965564\\
65.3408163265306	39.2492553013117\\
67.3795918367347	39.331877219243\\
69.4183673469388	39.1301661789571\\
71.4571428571429	39.4353269620942\\
73.4959183673469	39.1022786775216\\
75.534693877551	39.3233385124622\\
77.5734693877551	38.954004951421\\
79.6122448979592	38.9217253930306\\
81.6510204081633	39.2795151183426\\
83.6897959183673	39.0131711129328\\
85.7285714285714	38.9880611213865\\
87.7673469387755	38.7203531978553\\
89.8061224489796	38.7821785546986\\
91.8448979591837	38.6618303481765\\
93.8836734693878	38.534636619911\\
95.9224489795918	38.6460585212546\\
97.9612244897959	38.4186261101685\\
100	38.2055634393474\\
};

  \addplot [color=white, draw=none, mark=x, thick, mark size=6pt,    mark repeat=4, mark options={ black}]
  table[row sep=crcr]{%
0.1	15.7826945564707\\
2.13877551020408	12.8227794694421\\
4.17755102040816	17.3322825723413\\
6.21632653061225	20.7529706781795\\
8.25510204081633	23.162914350919\\
10.2938775510204	25.5162423136038\\
12.3326530612245	27.1684251691454\\
14.3714285714286	28.8028664880117\\
16.4102040816327	30.0528605463547\\
18.4489795918367	31.112079869195\\
20.4877551020408	32.365708075359\\
22.5265306122449	33.2991535587271\\
24.565306122449	34.0993906018257\\
26.6040816326531	34.9633739676122\\
28.6428571428571	35.722006668941\\
30.6816326530612	36.2134549827439\\
32.7204081632653	36.741859901664\\
34.7591836734694	37.3410583904502\\
36.7979591836735	37.769758666727\\
38.8367346938776	38.2360696736524\\
40.8755102040816	38.527440978107\\
42.9142857142857	39.14801152301\\
44.9530612244898	39.4748621146406\\
46.9918367346939	39.6204863080031\\
49.030612244898	39.8421670423755\\
51.069387755102	40.1460328490913\\
53.1081632653061	40.6192331718563\\
55.1469387755102	40.4979288291862\\
57.1857142857143	40.856934763018\\
59.2244897959184	40.8968962981578\\
61.2632653061224	40.8645931282344\\
63.3020408163265	40.9787966850976\\
65.3408163265306	41.1840689724109\\
67.3795918367347	41.4049186806298\\
69.4183673469388	41.3216460790512\\
71.4571428571429	41.5802469660914\\
73.4959183673469	41.353404646231\\
75.534693877551	41.5502695970425\\
77.5734693877551	41.3477323080137\\
79.6122448979592	41.341454075785\\
81.6510204081633	41.7909718986606\\
83.6897959183673	41.491496990197\\
85.7285714285714	41.5701985525483\\
87.7673469387755	41.3824449706613\\
89.8061224489796	41.4394225883865\\
91.8448979591837	41.3932701471288\\
93.8836734693878	41.3317490466252\\
95.9224489795918	41.4514439432322\\
97.9612244897959	41.2487938851034\\
100	41.1022511502342\\
};

\end{axis}

\begin{axis}[%
width=5.833in,
height=4.375in,
at={(0in,0in)},
scale only axis,
xmin=0,
xmax=1,
ymin=0,
ymax=1,
axis line style={draw=none},
ticks=none,
axis x line*=bottom,
axis y line*=left
]
\end{axis}
\end{tikzpicture}%

%% file: pow_comp.tex
%
%
\definecolor{mycolor1}{rgb}{0.00000,0.44700,0.74100}%
\definecolor{mycolor2}{rgb}{0.85000,0.32500,0.09800}%
\definecolor{mycolor3}{rgb}{0.92900,0.69400,0.12500}%
\definecolor{mycolor4}{rgb}{0.49400,0.18400,0.55600}%
\definecolor{mycolor5}{rgb}{0.46600,0.67400,0.18800}%
\definecolor{mycolor6}{rgb}{0.30100,0.74500,0.93300}%
\begin{tikzpicture}[scale=0.33, transform shape,font=\LARGE]

\begin{axis}[%
width=4.521in,
height=3.566in,
at={(0.758in,0.481in)},
scale only axis,
xmin=0,
xmax=1,
xlabel style={font=\huge\color{white!15!black}},
xlabel={Energy split factor $\gamma$},
ymin=0,
ymax=70,
ylabel style={font=\huge\color{white!15!black}},
ylabel={Average Rate [nats/s/Hz]},
axis background/.style={fill=white},
xmajorgrids,
ymajorgrids,
legend style={at={(0.294,0)}, anchor=south west, legend cell align=left, align=left, draw=white!15!black,font=\huge,
inner sep=1pt,
row sep=0.5pt}
]
 \addplot [
            color=mycolor1, 
            line width=2pt,
            mark=o, 
            mark options={solid, mycolor1},
            mark repeat=4 
        ] 
  table[row sep=crcr]{%
0	0\\
0.0101010101010101	9.78921458688616\\
0.0202020202020202	12.1443666043192\\
0.0303030303030303	13.6663084318453\\
0.0404040404040404	14.8096023326788\\
0.0505050505050505	15.7328875589059\\
0.0606060606060606	16.511466798177\\
0.0707070707070707	17.1873334361739\\
0.0808080808080808	17.7864101641484\\
0.0909090909090909	18.325869706731\\
0.101010101010101	18.8176973464694\\
0.111111111111111	19.27060121517\\
0.121212121212121	19.6911143019711\\
0.131313131313131	20.0842676362949\\
0.141414141414141	20.4540209417121\\
0.151515151515152	20.8035488266905\\
0.161616161616162	21.1354371501202\\
0.171717171717172	21.4518214759565\\
0.181818181818182	21.7544870185739\\
0.191919191919192	22.0449422851066\\
0.202020202020202	22.3244743262648\\
0.212121212121212	22.5941908590577\\
0.222222222222222	22.8550528448953\\
0.232323232323232	23.1079000134809\\
0.242424242424242	23.353471095486\\
0.252525252525253	23.5924200330004\\
0.262626262626263	23.8253290950715\\
0.272727272727273	24.0527195853647\\
0.282828282828283	24.2750606574167\\
0.292929292929293	24.4927766287478\\
0.303030303030303	24.7062530940301\\
0.313131313131313	24.9158420699351\\
0.323232323232323	25.1218663536074\\
0.333333333333333	25.3246232383151\\
0.343434343434343	25.5243877004696\\
0.353535353535354	25.7214151495587\\
0.363636363636364	25.9159438149323\\
0.373737373737374	26.1081968295867\\
0.383838383838384	26.2983840602207\\
0.393939393939394	26.4867037242061\\
0.404040404040404	26.6733438272328\\
0.414141414141414	26.8584834498719\\
0.424242424242424	27.0422939068598\\
0.434343434343434	27.2249397993265\\
0.444444444444444	27.406579977295\\
0.454545454545455	27.5873684274439\\
0.464646464646465	27.7674550992422\\
0.474747474747475	27.9469866810605\\
0.484848484848485	28.1261073366713\\
0.494949494949495	28.3049594116349\\
0.505050505050505	28.4836841183762\\
0.515151515151515	28.662422208284\\
0.525252525252525	28.8413146388767\\
0.535353535353535	29.0205032439721\\
0.545454545454545	29.2001314148729\\
0.555555555555556	29.3803448008285\\
0.565656565656566	29.5612920374766\\
0.575757575757576	29.7431255126114\\
0.585858585858586	29.926002179502\\
0.595959595959596	30.1100844291206\\
0.606060606060606	30.2955410340854\\
0.616161616161616	30.4825481789318\\
0.626262626262626	30.6712905935763\\
0.636363636363636	30.8619628096221\\
0.646464646464647	31.0547705626106\\
0.656565656565657	31.2499323675933\\
0.666666666666667	31.4476813007154\\
0.676767676767677	31.6482670261206\\
0.686868686868687	31.8519581157861\\
0.696969696969697	32.0590447203374\\
0.707070707070707	32.2698416621222\\
0.717171717171717	32.4846920386892\\
0.727272727272727	32.7039714464734\\
0.737373737373737	32.9280929625232\\
0.747474747474748	33.1575130586824\\
0.757575757575758	33.3927386708213\\
0.767676767676768	33.6343357097861\\
0.777777777777778	33.8829393868512\\
0.787878787878788	34.1392668435331\\
0.797979797979798	34.4041327367509\\
0.808080808080808	34.6784686550691\\
0.818181818181818	34.9633475598686\\
0.828282828282828	35.2600149027892\\
0.838383838383838	35.5699287403703\\
0.848484848484849	35.8948121659547\\
0.858585858585859	36.2367229021244\\
0.868686868686869	36.5981472753004\\
0.878787878787879	36.9821296088968\\
0.888888888888889	37.3924543789157\\
0.898989898989899	37.8339092732082\\
0.909090909090909	38.3126765333827\\
0.919191919191919	38.8369358732287\\
0.929292929292929	39.4178331291776\\
0.939393939393939	40.0711182488347\\
0.94949494949495	40.8200984182794\\
0.95959595959596	41.7014218388707\\
0.96969696969697	42.7777455732357\\
0.97979797979798	44.1703404473849\\
0.98989898989899	46.1681678261326\\
1	50.2812580614532\\
};
\addlegendentry{$N_c$=1 Sens.}

 \addplot [
            color=mycolor2, 
            line width=2pt,
            mark=asterisk, 
            mark options={solid, mycolor2},
            mark repeat=4 
        ] 
  table[row sep=crcr]{%
0	0\\
0.0101010101010101	11.435375008587\\
0.0202020202020202	14.4075223702873\\
0.0303030303030303	16.3588500728917\\
0.0404040404040404	17.8392009185439\\
0.0505050505050505	19.0432322325038\\
0.0606060606060606	20.064222901666\\
0.0707070707070707	20.9545643827273\\
0.0808080808080808	21.7467848174167\\
0.0909090909090909	22.4625367179898\\
0.101010101010101	23.1169944898597\\
0.111111111111111	23.7212233418408\\
0.121212121212121	24.283551334889\\
0.131313131313131	24.8104104891594\\
0.141414141414141	25.3068765668232\\
0.151515151515152	25.7770288112839\\
0.161616161616162	26.2241974214869\\
0.171717171717172	26.6511384614182\\
0.181818181818182	27.0601604011142\\
0.191919191919192	27.4532175478372\\
0.202020202020202	27.8319802789097\\
0.212121212121212	28.1978886836565\\
0.222222222222222	28.5521941215527\\
0.232323232323232	28.8959918345023\\
0.242424242424242	29.2302468383849\\
0.252525252525253	29.5558146980758\\
0.262626262626263	29.873458359999\\
0.272727272727273	30.183861913312\\
0.282828282828283	30.4876419342035\\
0.292929292929293	30.7853569107475\\
0.303030303030303	31.0775151304597\\
0.313131313131313	31.3645813270505\\
0.323232323232323	31.6469823185583\\
0.333333333333333	31.9251118202633\\
0.343434343434343	32.1993345784455\\
0.353535353535354	32.4699899422165\\
0.363636363636364	32.737394968221\\
0.373737373737374	33.0018471354145\\
0.383838383838384	33.2636267332421\\
0.393939393939394	33.5229989755256\\
0.404040404040404	33.7802158835676\\
0.414141414141414	34.0355179749296\\
0.424242424242424	34.2891357886672\\
0.434343434343434	34.5412912732236\\
0.444444444444444	34.7921990594902\\
0.454545454545455	35.042067638558\\
0.464646464646465	35.2911004612885\\
0.474747474747475	35.5394969749231\\
0.484848484848485	35.7874536104485\\
0.494949494949495	36.035164733286\\
0.505050505050505	36.2828235690255\\
0.515151515151515	36.5306231153615\\
0.525252525252525	36.7787570510659\\
0.535353535353535	37.0274206527657\\
0.545454545454545	37.2768117304561\\
0.555555555555556	37.5271315930904\\
0.565656565656566	37.778586056263\\
0.575757575757576	38.0313865049505\\
0.585858585858586	38.2857510255557\\
0.595959595959596	38.5419056231356\\
0.606060606060606	38.8000855417689\\
0.616161616161616	39.0605367086049\\
0.626262626262626	39.3235173253401\\
0.636363636363636	39.5892996348379\\
0.646464646464647	39.8581718955087\\
0.656565656565657	40.1304406021417\\
0.666666666666667	40.4064329994166\\
0.676767676767677	40.6864999437214\\
0.686868686868687	40.9710191806652\\
0.696969696969697	41.2603991204875\\
0.707070707070707	41.555083212321\\
0.717171717171717	41.8555550421767\\
0.727272727272727	42.1623443102293\\
0.737373737373737	42.476033882717\\
0.747474747474748	42.7972681656359\\
0.757575757575758	43.1267631157187\\
0.767676767676768	43.4653182950311\\
0.777777777777778	43.8138314976292\\
0.787878787878788	44.1733166427214\\
0.797979797979798	44.5449258572624\\
0.808080808080808	44.9299769896269\\
0.818181818181818	45.3299882471517\\
0.828282828282828	45.7467222991853\\
0.838383838383838	46.1822431370077\\
0.848484848484849	46.6389903992246\\
0.858585858585859	47.1198780320211\\
0.868686868686869	47.6284275278301\\
0.878787878787879	48.1689513984278\\
0.888888888888889	48.7468114885557\\
0.898989898989899	49.3687920588413\\
0.909090909090909	50.0436548707612\\
0.919191919191919	50.7829944872708\\
0.929292929292929	51.6026126069658\\
0.939393939393939	52.5248424713274\\
0.94949494949495	53.582740428747\\
0.95959595959596	54.8282974709492\\
0.96969696969697	56.35043134683\\
0.97979797979798	58.3213240778408\\
0.98989898989899	61.1516560024758\\
1	67.0603709384875\\
};
\addlegendentry{$N_c$=6 Sens.}

 \addplot [
            color=mycolor3, 
            line width=2pt,
            mark=diamond, 
            mark options={solid, mycolor3},
            mark repeat=4 
        ] 
  table[row sep=crcr]{%
0	39.9022173032034\\
0.0101010101010101	39.8795495518848\\
0.0202020202020202	39.8566260812376\\
0.0303030303030303	39.8334414637408\\
0.0404040404040404	39.8099900992305\\
0.0505050505050505	39.7862662075261\\
0.0606060606060606	39.762263820659\\
0.0707070707070707	39.7379767746783\\
0.0808080808080808	39.7133987010051\\
0.0909090909090909	39.6885230173059\\
0.101010101010101	39.6633429178524\\
0.111111111111111	39.6378513633317\\
0.121212121212121	39.6120410700708\\
0.131313131313131	39.5859044986328\\
0.141414141414141	39.5594338417421\\
0.151515151515152	39.5326210114878\\
0.161616161616162	39.5054576257573\\
0.171717171717172	39.4779349938395\\
0.181818181818182	39.4500441011388\\
0.191919191919192	39.4217755929324\\
0.202020202020202	39.3931197570974\\
0.212121212121212	39.3640665057299\\
0.222222222222222	39.3346053555684\\
0.232323232323232	39.304725407128\\
0.242424242424242	39.2744153224416\\
0.252525252525253	39.2436633012953\\
0.262626262626263	39.2124570558347\\
0.272727272727273	39.1807837834054\\
0.282828282828283	39.148630137479\\
0.292929292929293	39.1159821965012\\
0.303030303030303	39.0828254304795\\
0.313131313131313	39.0491446651131\\
0.323232323232323	39.0149240432431\\
0.333333333333333	38.9801469833811\\
0.343434343434343	38.9447961350456\\
0.353535353535354	38.908853330607\\
0.363636363636364	38.8722995333096\\
0.373737373737374	38.8351147810996\\
0.383838383838384	38.7972781258489\\
0.393939393939394	38.7587675675116\\
0.404040404040404	38.7195599827013\\
0.414141414141414	38.6796310471105\\
0.424242424242424	38.6389551511256\\
0.434343434343434	38.5975053079087\\
0.444444444444444	38.5552530531257\\
0.454545454545455	38.5121683353932\\
0.464646464646465	38.4682193963952\\
0.474747474747475	38.4233726394805\\
0.484848484848485	38.3775924853851\\
0.494949494949495	38.3308412135402\\
0.505050505050505	38.2830787872025\\
0.515151515151515	38.2342626603879\\
0.525252525252525	38.1843475642932\\
0.535353535353535	38.1332852705329\\
0.545454545454545	38.0810243281102\\
0.555555555555556	38.0275097705478\\
0.565656565656566	37.9726827890274\\
0.575757575757576	37.9164803666953\\
0.585858585858586	37.8588348684687\\
0.595959595959596	37.7996735796906\\
0.606060606060606	37.7389181857896\\
0.616161616161616	37.6764841836661\\
0.626262626262626	37.6122802137755\\
0.636363636363636	37.5462072997459\\
0.646464646464647	37.4781579797511\\
0.656565656565657	37.4080153106231\\
0.666666666666667	37.3356517216846\\
0.676767676767677	37.2609276902682\\
0.686868686868687	37.1836902046061\\
0.696969696969697	37.1037709718163\\
0.707070707070707	37.020984318577\\
0.717171717171717	36.9351247190809\\
0.727272727272727	36.8459638680314\\
0.737373737373737	36.7532471944882\\
0.747474747474748	36.6566896834508\\
0.757575757575758	36.5559708336087\\
0.767676767676768	36.4507285279882\\
0.777777777777778	36.340551523937\\
0.787878787878788	36.2249701721406\\
0.797979797979798	36.1034448394077\\
0.808080808080808	35.975351318948\\
0.818181818181818	35.8399622371847\\
0.828282828282828	35.6964230642089\\
0.838383838383838	35.5437207354417\\
0.848484848484849	35.380641978553\\
0.858585858585859	35.205717014399\\
0.868686868686869	35.0171420170917\\
0.878787878787879	34.8126699478798\\
0.888888888888889	34.5894529356499\\
0.898989898989899	34.343807928952\\
0.909090909090909	34.0708560463163\\
0.919191919191919	33.7639442257541\\
0.929292929292929	33.413670126268\\
0.939393939393939	33.0061323489406\\
0.94949494949495	32.5195291514262\\
0.95959595959596	31.9168020680124\\
0.96969696969697	31.127131102101\\
0.97979797979798	29.9870731389339\\
0.98989898989899	27.9499647679732\\
1	0\\
};
\addlegendentry{Comm.}

 \addplot [
            color=mycolor6, 
            line width=2pt,
            mark=+, 
            mark options={solid, mycolor6},
            mark repeat=4 
        ]
  table[row sep=crcr]{%
0	40.0423782120413\\
0.0101010101010101	40.0423782120413\\
0.0202020202020202	40.0423782120413\\
0.0303030303030303	40.0423782120413\\
0.0404040404040404	40.0423782120413\\
0.0505050505050505	40.0423782120413\\
0.0606060606060606	40.0423782120413\\
0.0707070707070707	40.0423782120413\\
0.0808080808080808	40.0423782120413\\
0.0909090909090909	40.0423782120413\\
0.101010101010101	40.0423782120413\\
0.111111111111111	40.0423782120413\\
0.121212121212121	40.0423782120413\\
0.131313131313131	40.0423782120413\\
0.141414141414141	40.0423782120413\\
0.151515151515152	40.0423782120413\\
0.161616161616162	40.0423782120413\\
0.171717171717172	40.0423782120413\\
0.181818181818182	40.0423782120413\\
0.191919191919192	40.0423782120413\\
0.202020202020202	40.0423782120413\\
0.212121212121212	40.0423782120413\\
0.222222222222222	40.0423782120413\\
0.232323232323232	40.0423782120413\\
0.242424242424242	40.0423782120413\\
0.252525252525253	40.0423782120413\\
0.262626262626263	40.0423782120413\\
0.272727272727273	40.0423782120413\\
0.282828282828283	40.0423782120413\\
0.292929292929293	40.0423782120413\\
0.303030303030303	40.0423782120413\\
0.313131313131313	40.0423782120413\\
0.323232323232323	40.0423782120413\\
0.333333333333333	40.0423782120413\\
0.343434343434343	40.0423782120413\\
0.353535353535354	40.0423782120413\\
0.363636363636364	40.0423782120413\\
0.373737373737374	40.0423782120413\\
0.383838383838384	40.0423782120413\\
0.393939393939394	40.0423782120413\\
0.404040404040404	40.0423782120413\\
0.414141414141414	40.0423782120413\\
0.424242424242424	40.0423782120413\\
0.434343434343434	40.0423782120413\\
0.444444444444444	40.0423782120413\\
0.454545454545455	40.0423782120413\\
0.464646464646465	40.0423782120413\\
0.474747474747475	40.0423782120413\\
0.484848484848485	40.0423782120413\\
0.494949494949495	40.0423782120413\\
0.505050505050505	40.0423782120413\\
0.515151515151515	40.0423782120413\\
0.525252525252525	40.0423782120413\\
0.535353535353535	40.0423782120413\\
0.545454545454545	40.0423782120413\\
0.555555555555556	40.0423782120413\\
0.565656565656566	40.0423782120413\\
0.575757575757576	40.0423782120413\\
0.585858585858586	40.0423782120413\\
0.595959595959596	40.0423782120413\\
0.606060606060606	40.0423782120413\\
0.616161616161616	40.0423782120413\\
0.626262626262626	40.0423782120413\\
0.636363636363636	40.0423782120413\\
0.646464646464647	40.0423782120413\\
0.656565656565657	40.0423782120413\\
0.666666666666667	40.0423782120413\\
0.676767676767677	40.0423782120413\\
0.686868686868687	40.0423782120413\\
0.696969696969697	40.0423782120413\\
0.707070707070707	40.0423782120413\\
0.717171717171717	40.0423782120413\\
0.727272727272727	40.0423782120413\\
0.737373737373737	40.0423782120413\\
0.747474747474748	40.0423782120413\\
0.757575757575758	40.0423782120413\\
0.767676767676768	40.0423782120413\\
0.777777777777778	40.0423782120413\\
0.787878787878788	40.0423782120413\\
0.797979797979798	40.0423782120413\\
0.808080808080808	40.0423782120413\\
0.818181818181818	40.0423782120413\\
0.828282828282828	40.0423782120413\\
0.838383838383838	40.0423782120413\\
0.848484848484849	40.0423782120413\\
0.858585858585859	40.0423782120413\\
0.868686868686869	40.0423782120413\\
0.878787878787879	40.0423782120413\\
0.888888888888889	40.0423782120413\\
0.898989898989899	40.0423782120413\\
0.909090909090909	40.0423782120413\\
0.919191919191919	40.0423782120413\\
0.929292929292929	40.0423782120413\\
0.939393939393939	40.0423782120413\\
0.94949494949495	40.0423782120413\\
0.95959595959596	40.0423782120413\\
0.96969696969697	40.0423782120413\\
0.97979797979798	40.0423782120413\\
0.98989898989899	40.0423782120413\\
1	40.0423782120413\\
};
\addlegendentry{Comm.-only network}







\end{axis}

\end{tikzpicture}%